\documentclass[twocolumn,runningheads]{svjour2}
\smartqed  % flush right qed marks, e.g. at end of proof
\usepackage{graphicx}
\journalname{Astrophysics and Space Science}
\begin{document}

\title{Synchrotron emission in the fast cooling
regime: which spectra can be explained?}

\author{Evgeny V. Derishev}

\institute{E.V. Derishev \at Institute of Applied Physics, 46
Ulyanov st., 603950 Nizhny Novgorod, Russia}

\date{Received: date / Accepted: date}
% The correct dates will be entered by the editor

\maketitle

\begin{abstract}
We consider the synchrotron emission from relativistic shocks
assuming that the radiating electrons cool rapidly (either through
synchrotron or any other radiation mechanism). It is shown that
the theory of synchrotron emission in the fast cooling regime can
account for a wide range of spectral shapes. In particular, the
magnetic field, which decays behind the shock front, brings enough
flexibility to the theory to explain the majority of gamma-ray
burst spectra even in the parameter-free fast cooling regime.
Also, we discuss whether location of the peak in observed spectral
energy distributions of gamma-ray bursts and active galactic
nuclei can be made consistent with predictions of diffusive shock
acceleration theory, and find that the answer is negative. This
result is a strong indication that a particle injection mechanism,
other than the standard shock acceleration, works in relativistic
shocks.
\end{abstract}

\section{Introduction}

The synchrotron radiation is a very common emission mechanism
among various astrophysical sources: supernova remnants, active
galactic nuclei (AGNs), gamma-ray bursts (GRBs), etc. It allows to
explain a wide range of different broad-band spectra by adjusting
distribution functions of radiating particles. The models
employing synchrotron radiation are more constrained in the
so-called fast cooling regime, where the bulk of particles radiate
away their energy before escaping the emitting region or loosing
energy through adiabatic cooling. The fast cooling regime is
desirable for very luminous objects like AGNs and GRBs since the
energy limitations of their central engines imply rather high
radiative efficiency. Also, this regime is what one expects in
these objects theoretically, based on the standard assumption that
the magnetic field strength is close to the equipartition value,
which is enough to ensure fast cooling even through synchrotron
radiation alone.

In this paper we put emphasis on AGNs and GRBs, investigating
general characteristics of their spectral energy distributions
(SEDs), such as location of the peak and the spectral indices
below and above the peak. We analyze whether the observed spectra
can be made consistent with the standard particle acceleration
models or in principle explained by the synchrotron radiation
(including into consideration models with non-uniform magnetic
field distribution).

\section{Relation of particle injection to the observed spectra}

In the generally accepted model, the observed emission of AGNs and
GRBs comes from a succession of multiple mildly relativistic
shocks, which form within continuous outflow of magnetized plasma
with the bulk Lorentz factor $\Gamma_b \gg 1$. These internal
shocks accelerate electrons, which are eventually advected
downstream of the shock front, where they produce synchrotron
radiation.

The schematic distribution of electrons injected at the shock
front, $f(\gamma)$ (where $\gamma$ is the Lorentz factor of an
electron), is presented in Fig.~1. Most of the electrons belong to
"thermal" population with average energy of the order of the
average energy of shocked protons $\Gamma m_p c^2$ ($\Gamma$ is
the shock Lorentz factor). At a certain (low) level there is a
smooth transition from the "thermal" distribution to a power-law
non-thermal one formed by shock-accelerated electrons. The
power-law cuts off at an energy where the radiative losses start
to prevail over the acceleration energy gain.

\begin{figure}
\centering
  \includegraphics[width=0.9\columnwidth]{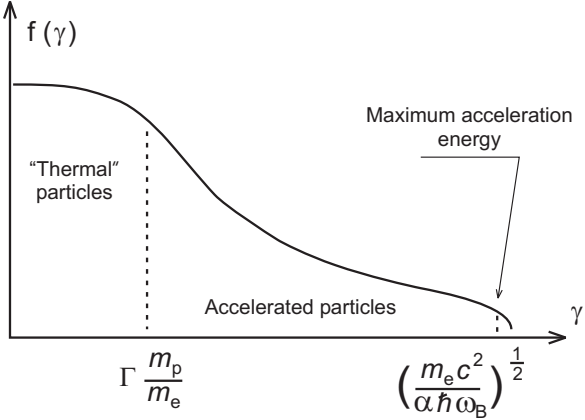}
\caption{Distribution of particles injected at the shock front as
a result of diffusive shock acceleration. The cut-off energy is
defined assuming that the synchrotron radiation is the only energy
loss mechanism. Here $\Gamma$ is the shock Lorentz factor,
$\omega_{\rm B} = eB/m_e c$ -- electron's gyrofrequency, and
$\alpha$ -- the fine structure constant.}
\label{injection}
\end{figure}

In the fast cooling regime, the electrons' distribution function
changes as the electrons are advected away from the shock front
and cool: the cut-off shifts towards progressively smaller
energies. However, the observed luminosity is the integral of
emissivity along the line of sight, so that in the one-zone model
with constant magnetic-field strength downstream one only needs to
know the integral of the distribution function along the shock
normal, $N(\gamma)$, to calculate the observed spectrum (due to
the effect of relativistic beaming, only those portions of the
shock whose normal is closely aligned with the line of sight
contribute to the observed emission). In effect, the problem is
reduced to the case of uniform injection.

The electron distribution function can be found from the
continuity equation in the energy-momentum space
\begin{equation}
\frac{\partial N}{\partial t}+{\rm div}\, (\dot{\gamma}N) =
f(\gamma),
\end{equation}
which gives stationary solution
\begin{equation}\label{cool-distr}
N(\gamma) = - \frac{1}{\dot{\gamma}} \int_{\gamma}^{\infty}
f(\gamma^{\prime})\, d \gamma^{\prime}.
\end{equation}
The corresponding SED (assuming the radiation is mono\-chromatic
with frequency $\nu \propto \gamma^2$) is:
\begin{equation}\label{cool-SED}
\nu F_{\nu} \propto \frac{d F}{d \ln \gamma} \propto \eta \gamma
\int_{\gamma}^{\infty} f(\gamma^{\prime})\, d \gamma^{\prime},
\end{equation}
where $\eta (\gamma)$ is the fraction of electron's energy
transferred to the synchrotron radiation.

Apparently, there are two prominent features in the distribution
function given by Eq. \ref{cool-distr}, and the peak in an
observed SED can be related to either of them. One, more or less
standard assignment (see Fig. 2), links the peak to the break at
the transitional region between "thermal" and non-thermal
electrons. Alternatively, one may link the peak to the cut-off
region of the electron distribution (see Fig. 3).

\begin{figure}
\centering
  \includegraphics[width=0.9\columnwidth]{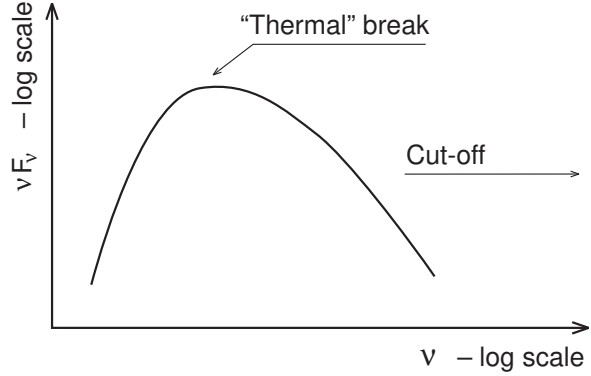}
\caption{Standard assignment of spectral features.}
\label{standard}
\end{figure}

\begin{figure}
\centering
  \includegraphics[width=0.9\columnwidth]{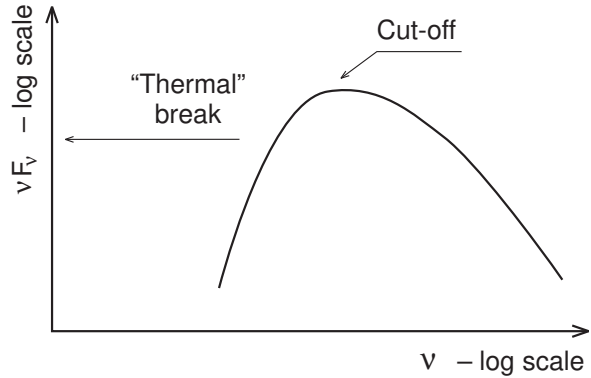}
\caption{Alternative assignment of spectral features.}
\label{altern}
\end{figure}

The alternative assignment of the SED peak has more explanatory
power as far as spectral index above the peak is considered. The
predicted cut-off shape for a particle distribution originating
from acceleration at a relativistic shock is model-dependent and
very different from a simple exponential cut-off \cite{cutoff}, in
accordance with the observed spectral indices, which are indeed
different for different sources. On the contrary, the standard
assignment of the SED peak has difficulty in explaining this
difference since diffusive shock acceleration gives a universal
power-law $f(\gamma)\propto \gamma^{-2.2}$ (e.g.,
\cite{Waxm})\footnote{It has been claimed in a number of recent
papers \cite{NO,Pell}, that more realistic models of particle
scattering lead to softer and model-dependent spectra of
accelerated particles. On the other hand, the acceleration is
inefficient (or even absent) in these models. Thus, wherever
relativistic shocks efficiently accelerate electrons, the injected
particle distribution must be close to $f(\gamma)\propto
\gamma^{-2.2}$.}, resulting in very hard ($\nu F_{\nu} \propto
\nu^{-0.1}$) and {\em universal} spectra above the peak.

\section{Location of the peak}

In the internal shock model, the comoving-frame Lorentz factor of
thermal electrons is $\gamma \sim m_p/m_e$, so that the standard
assignment of SED peak implies that it is observed at the energy
\begin{equation}\label{thermal-peak}
\varepsilon_{\rm peak} \sim \Gamma_b \left( \frac{m_p}{m_e}
\right)^2 \frac{\hbar e B}{m_e c},
\end{equation}
where $B$ is the magnetic field strength in the comoving frame.

The internal shocks form at a distance $\sim \Gamma_b^2 t_v c$
from the central engine, where $t_v$ is the source's variability
timescale, which is related to the size of central engine.
Assuming equipartition between the magnetic-field and the
radiation-field energy densities, we find that the magnetic field
strength is
\begin{equation}\label{B}
B \sim \frac{L^{1/2}}{\Gamma_b^3 t_v c^{3/2}},
\end{equation}
where $L$ is the apparent luminosity of a source. Substituting Eq.
\ref{B} into Eq. \ref{thermal-peak}, we get the location of SED
peak:
\begin{equation}\label{SEDpeak1}
\varepsilon_{\rm peak} \sim \left( \frac{m_p}{m_e} \right)^2
\frac{\hbar e L^{1/2}}{\Gamma_b^2 t_v m_e c^{5/2}} \sim
\frac{L_{51}^{1/2}}{\Gamma_3^2 t_{-3}}\, {\rm MeV}.
\end{equation}
Here $L_{51}$ is the luminosity in units $10^{51}$ erg/s, $t_{-3}$
the variability timescale in units $10^{-3}$ s, and $\Gamma_3$ the
bulk Lorentz factor in units $10^3$. These units are chosen
because they are standard parameters of a typical gamma-ray burst.
Equation \ref{SEDpeak1} predicts that GRBs have their spectral
energy distributions peaked at roughly 1~MeV, in accordance with
observations. However, a typical AGN with luminosity $L \sim
10^{45}$ erg/s, variability timescale $t_v \sim 10^4$~s, and the
bulk Lorentz factor $\Gamma_b \sim 10$ should have an SED peaked
at around 1~eV -- hardly enough to explain IR-peaked AGNs, and far
too low for MeV-peaked blazars.

Assigning the SED peak to the cut-off region of the electron
injection function poses difficulties as well. Indeed, in the case
of Bohm diffusion the acceleration rate at a relativistic shock is
$m_e c^2 \dot{\gamma} \simeq eBc$ and hence the maximum
acceleration energy for electrons, determined from the balance
between energy gain and radiative losses,
\begin{equation}
m_e c^2 \dot{\gamma} = -  \frac{1}{\eta (\gamma)} \frac{4}{9}
\left( \frac{e^2}{m_e c^2} \right)^2 \gamma^2 B^2 c\, ,
\end{equation}
equals to
\begin{equation}
\gamma_{\rm max} \simeq \frac{3}{2} \frac{\left[\eta(\gamma_{\rm
max})\right]^{1/2} m_e c^2}{\sqrt{e^3 B}}\, ,
\end{equation}
and the associated SED peak of their synchrotron emission is at
\begin{equation}\label{SEDpeak2}
\varepsilon_{\rm peak} \simeq \Gamma_b \gamma_{\rm max}^2
\frac{\hbar e B}{m_e c} \simeq \frac{9}{4} \Gamma_b
\eta(\gamma_{\rm max}) \frac{m_e c^2}{\alpha_f} ,
\end{equation}
where $\alpha_f$ is the fine structure constant. The location of
SED peak given by Eq. \ref{SEDpeak2} does not explicitly depend on
the magnetic field strength (although depends on it implicitly
through $\gamma_{\rm max}(B)$ and then $\eta(\gamma_{\rm max})$)
and even for moderate values of the bulk Lorentz factor appears to
be in the GeV range, unless the synchrotron efficiency $\eta$ is
unreasonably low.

However, a diffusion faster than the Bohm one results in a smaller
peak energy, in better agreement with observations. For example,
if the plasma in the bulk outflow can sustain magnetic-field
inhomogeneities with sizes $\geq \ell_{\rm c}$, then the electron
scattering length can be made no larger than
\begin{equation}\label{diff-lim}
\ell_{\rm s} = \ell_{\rm c} \left( \frac{r_{g}}{\ell_{\rm c}}
\right)^2 = \frac{\left( \gamma m_e c^2 \right)^2}{\ell_{\rm c}
e^2 B^2},
\end{equation}
where $r_{g} = \gamma m_e c^2/eB$ is the electron's gyroradius.
The value of the scattering length given by the above equation
minimizes acceleration rate, which becomes equal to
\begin{equation}\label{min-accel}
\dot{\gamma} \simeq \frac{1}{\gamma} \frac{\ell_{\rm c} e^2 B^2
c}{\left(m_e c^2 \right)^2}.
\end{equation}
Radiative losses terminate acceleration at
\begin{equation}
\gamma_{\rm max} \simeq \left( \eta(\gamma_{\rm max})
\frac{\ell_{\rm c}}{r_e} \right)^{1/3},
\end{equation}
where $r_e$ is the classical electron radius. Consequently, the
peak of synchrotron SED is at
\begin{equation}\label{SEDpeak3}
\varepsilon_{\rm peak} \simeq  \Gamma_b \left( \eta(\gamma_{\rm
max})\, \frac{\ell_{\rm c}}{r_{\rm g0}} \right)^{2/3} \left(
\frac{\alpha_f B}{B_{\rm cr}} \right)^{1/3} \frac{m_e
c^2}{\alpha_f},
\end{equation}
where $B_{\rm cr} \simeq 4.4 \times 10^{13}$~G is the Schwinger
magnetic field and $r_{\rm g0} = m_e c^2/eB$ the "cold"
gyroradius.

Pushing theory to the limits given by Eqs. \ref{diff-lim} and
\ref{min-accel} requires that electron's motion is random
small-angle scattering, that implies the magnetic field is
effectively uncorrelated on scales larger than $\ell_{\rm c}$ or
-- more precisely -- that the power spectrum of the magnetic field
$B_k^2$ peaks at $k=1/\ell_{\rm c}$ and decreases towards smaller
wavenumbers faster than $B_k^2 \propto k$. Under such
circumstances, the electrons emit mostly due to their interaction
with small-scale magnetic field inhomogeneities. If $\ell_{\rm c}
< r_{\rm g0}$, then electrons radiate in the undulator regime
\cite{ToptFleish,Medv} and the typical frequency of their emission
increases with decreasing magnetic-field scale.

Thus, the factor $\ell_{\rm c}/r_{\rm g0}$ in Eq. \ref{SEDpeak3}
can be made as small as unity, and prevalence of inverse-Compton
radiative losses ($\eta \ll 1$) further decreases
$\varepsilon_{\rm peak}$. In the case of GRBs, where $B \sim 10^5
- 10^6$~G, we find from Eq.~(\ref{SEDpeak3}) that the peak of
synchrotron SED can be located at just few MeV, roughly in
agreement with observations. For AGNs, whose typical value of the
magnetic field strength is $\sim 0.1$~G, it is not possible to
push the location of SED peak significantly below 1 keV while
keeping the synchrotron efficiency at an acceptable level. This
agrees with observations for many blazars, but cannot explain
IR-peaked AGNs.

\section{Low-frequency spectral index}

One of the problems in the interpretation of GRB emission as the
synchrotron radiation is that the low-frequency spectral index in
the fast-cooling regime is too soft. The hardest possible
injection $f(\gamma) = \delta (\gamma - \gamma_0)$ gives (see Eq.
\ref{cool-SED}) $\nu F_{\nu}  \propto \gamma \eta$ below the SED
peak (whose position in this case corresponds to $\gamma_0$), that
is $\nu F_{\nu} \propto \nu^{1/2}$ if the synchrotron efficiency
$\eta$ is constant.

However, in the synchrotron-self-Compton (SSC) model the
synchrotron efficiency is, generally speaking, a rising function
of the electron Lorentz factor. Due to the Klein-Nishina effect,
only photons whose frequency is less than $m_e c^2/h \gamma$ can
significantly contribute to the effective energy density of seed
radiation:
\begin{equation}
w_{\rm ph} (\gamma) \simeq \int_{0}^{\frac{m_e c^2}{h \gamma}}
w_{\nu}\, d\nu\, .
\end{equation}
Therefore, relative weight of synchrotron energy losses,
$\dot{\gamma} \propto -\gamma^2 B^2$, compared to the
inverse-Compton energy losses, $\dot{\gamma} \propto -\gamma^2
w_{\rm ph}$, increases for electrons with larger Lorentz factors.
In a consistent SSC model with prevalence of inverse Compton
radiative losses in the Klein-Nishina regime, the synchrotron
efficiency can rise as fast as $\eta(\gamma) \propto \gamma$,
leading to a rather hard low-frequency spectrum of synchrotron
radiation $\nu F_{\nu} \propto \nu$ \cite{ours}. This can be a
remedy for the synchrotron model of GRB emission for the majority
of bursts\footnote{A small number of bursts has still harder
low-frequency spectral index \cite{Preece}, and may require
employment of undulator emission in addition to the synchrotron
emission \cite{Medv}.}, although at a price of decreased
synchrotron efficiency.

In the internal shock model of GRBs applicability of this recipe
is limited by the fact, that the comptonization proceeds not very
deep in the Klein-Nishina regime. Moreover, in some bursts the
bulk of radiating electrons comptonize their own synchrotron
radiation in the Thomson regime. In this paper we propose another
way to obtain hard low-frequency spectra, which is also suitable
for SSC models in the Thomson regime. To reach this goal we note,
that the magnetic field behind the front of a relativistic shock
is produced by various plasma instabilities rather than by simple
MHD compression. This field is not frozen-in and must decay as the
shocked plasma moves away from the shock front. It means that the
different parts of the electron distribution function, which was
treated as a single integrated distribution earlier in this paper,
in fact "feel" different magnetic field strength (see Fig. 4),
whereas the spectrum and energy density of seed photons do not
significantly change inside a thin slab occupied by radiating
electrons.

\begin{figure}
\centering
  \includegraphics[width=0.9\columnwidth]{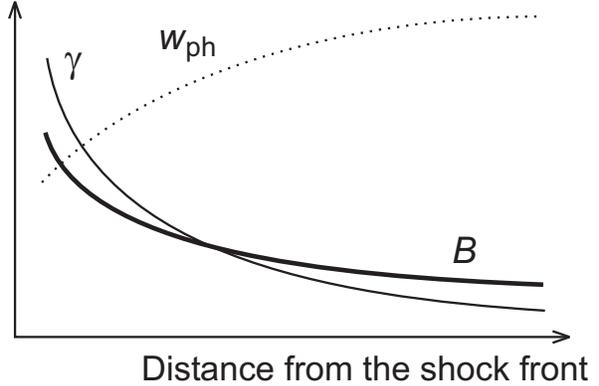}
\caption{The Lorentz factor of cooling electrons $\gamma$ (thin
line), the magnetic field strength $B$ (thick line), and the
effective energy density of seed photons $w_{\rm ph}$ (dotted
line), as functions of distance from the shock front.}
\label{B-decay}
\end{figure}

To investigate the main features of the proposed model, we start
with delta-functional injection  $f(\gamma) = \delta (\gamma -
\gamma_0)$ at the shock front. While cooling, the electrons are
advected downstream with constant velocity $v=c/3$, so that the
spacial derivative of their Lorentz factor is related to the time
derivative:
\begin{equation}
\frac{d\gamma}{dr} = \frac{3}{c} \frac{\partial \gamma}{\partial
t} = - 4 \gamma^2 \sigma_{\rm T} \left( w_{\rm ph}+
\frac{B^2}{8\pi} \right),
\end{equation}
where $r$ is the distance from the shock front and $\sigma_{\rm
T}$ the Thomson cross-section.

Let us assume that the spectrum of seed radiation is a power-law
$w_{\nu} \propto \nu^q$, where $-1<q<0$, so that $w_{\rm ph}
(\gamma) \propto \gamma^{-1-q}$. Then, the solution to the
continuity equation (Eq. \ref{cool-distr}) gives
\begin{equation}
N (\gamma) \propto \gamma^{q-1} \quad \mbox{for} \quad
\gamma<\gamma_0.
\end{equation}

In the case of low synchrotron efficiency $\eta \ll 1$ we obtain
\begin{equation}
\frac{d\gamma}{dr} \propto -\gamma^{1-q} \qquad \Rightarrow \quad
\gamma = \left( \frac{r}{r_0} \right)^{\frac{1}{q}} \quad
\mbox{for}  \quad \gamma \ll \gamma_0.
\end{equation}
Here $r_0$ is the radiation length of an electron with the Lorentz
factor $\gamma_0$. Assuming that the magnetic field strength is a
power-law function of the distance from the shock front, $B
\propto r^{-y}$, we find the synchrotron efficiency
\begin{equation}\label{eta}
\eta(\gamma) = \frac{B^2}{8\pi w_{\rm ph}} \propto r^{-2y}
\gamma^{1+q} \propto \gamma^{1+q-2qy}
\end{equation}
and the dependence of typical synchrotron frequency on the Lorentz
factor of radiating electrons
\begin{equation}\label{nu}
\nu(\gamma) \propto \gamma^2 B \propto \gamma^{2-qy}.
\end{equation}

Substituting Eqs. \ref{eta} and \ref{nu} into the right-hand-side
of Eq. \ref{cool-SED}, we find the emerging spectrum:
\begin{equation}\label{non-uniform-SED}
\nu F_{\nu} \propto \gamma \eta \propto
\nu^{\frac{2+q-2qy}{2-qy}}.
\end{equation}

There are two cases, which deserve particular attention. First of
all, choosing $q=-1$ in Eq. \ref{non-uniform-SED} models
comptonization in the Thomson regime: the effective energy density
of seed photons very weakly (logarithmically) depends on the
electron Lorentz factor and this dependence can be ignored when
calculating power-law indices. In this case we get
\begin{equation}
\nu F_{\nu} \propto \nu^{\frac{1+2y}{2+y}}.
\end{equation}
In principle, this model allows for spectra as hard as $\nu
F_{\nu} \propto \nu^2$, which is even harder than the
low-frequency asymptotic in the synchrotron spectrum of an
individual electron. However, the low-frequency spectrum remains
softer than $\nu F_{\nu} \propto \nu$, unless $y \geq 1$.

The self-consistent SSC model with comptonization in the
Klein-Nishina regime and $q=0$ eventually brings no difference
from its one-zone counterpart with constant magnetic field
strength:
\begin{equation}
\nu F_{\nu} \propto \nu\, .
\end{equation}

The above model is easy to generalize for the case of purely
synchrotron radiation ($\eta=1$) in non-uniform magnetic field.
The resulting spectrum is
\begin{equation}
\nu F_{\nu} \propto \nu^{\frac{1-2y}{2-3y}},
\end{equation}
which gives the familiar $\nu F_{\nu} \propto \nu^{1/2}$ outcome
for the constant magnetic field strength, and even softer spectra
for any decaying magnetic field. However, this case may be
interesting if for some reason the magnetic field strength
increases behind the shock front, so that $y$ is negative, -- such
a situation results in relatively hard low-frequency spectra.

\section{Conclusion}

We show that the low-energy (below the SED peak) part of AGN and
GRB spectra can be adequately described by synchrotron emission
models, which allow for a broad range of spectral indices. The
low-energy synchrotron spectra can be as hard as $\nu F_{\nu}
\propto \nu$ for the synchrotron-self-Compton model with
comptonization in the Klein-Nishina regime. Even harder spectra
are possible in the case, where the magnetic field decays behind
the shock front, and this result holds true even if comptonization
proceeds in the Thomson regime. However, this flexibility of
theory always comes at a price of low synchrotron efficiency.

Comparison of observed high-energy spectra above the SED peak to
the predictions of diffusive shock acceleration theory favors
assignment of the peak to the cut-off in the injected electron
distribution, whereas the standard assignment of the SED peak to
the thermal break produces too hard and invariable high-energy
spectral index.

Both assignments are consistent with the SED peak location
observed in GRBs, although the agreement is only marginal and may
be merely a chance coincidence. Furthermore, none of them can
explain the position of SED peaks in all of the observed AGNs or
even in the majority of them. We consider this result as a strong
indication that a particle injection mechanism, other than the
standard diffusive shock acceleration, is at work in relativistic
shocks.

\begin{acknowledgements}
E.V. Derishev acknowledges the support from the President of the
Russian Federation Program for Support of Young Scientists (grant
no. MK-2752.2005.2). This work was also supported by the RFBR
grants no. 05-02-17525 and 04-02-16987, the President of the
Russian Federation Program for Support of Leading Scientific
Schools (grant no. NSh-4588.2006.2), and the program "Origin and
Evolution of Stars and Galaxies" of the Presidium of the Russian
Academy of Science. This work was partly done during a visit to
the Max-Planck-Institut f\"ur Kernphysik.
\end{acknowledgements}

\end{document}